**Title:**

Surface plasmons on a Au waveguide electrode open new redox channels associated with the transfer of energetic carriers

**One sentence summary:**

Surface plasmons on a working electrode open optically controlled non-equilibrium redox channels associated with energetic carrier transfer to the redox species.

**Authors:**

Zohreh Hirbodvash,[1,2] Oleksiy Krupin,[2] Howard Northfield,[2] Anthony Olivieri,[2] Elena A. Baranova,[3,4] and Pierre Berini[1,2,5,*]

**Affiliations:**

[1]Dept. of Physics, University of Ottawa, 150 Louis Pasteur, Ottawa, Ontario, K1N 6N5, Canada
[2]Center for research in Photonics, University of Ottawa, 25 Templeton St., Ottawa, Ontario, K1N 6N5, Canada
[3]Department of Chemical and Biological Engineering, University of Ottawa, 161 Louis-Pasteur, Ottawa, ON K1N 6N5, Canada
[4]Centre for Catalysis Research and Innovation, University of Ottawa, 161 Louis-Pasteur, Ottawa, ON K1N 6N5, Canada
[5]School of Electrical Engineering and Computer Science, University of Ottawa, 800 King Edward Ave., Ottawa, Ontario, K1N 6N5, Canada
*Corresponding Author; berini@eecs.uottawa.ca

**Abstract:**

Plasmonic catalysis holds significant promise for opening new reaction pathways inaccessible thermally, or for improving the efficiency of chemical processes. However, challenges persist in distinguishing photothermal effects from effects due to energetic electrons and/or holes excited in the metal comprising a plasmonic structure. Here we use a Au stripe waveguide along which surface plasmon polaritons (SPPs) propagate, operating simultaneously as a working electrode in a 3-electrode electrochemical cell. Cyclic voltammograms were obtained under SPP excitation as a function of incident optical power and wavelength ($\lambda_0$ ~1350 nm), enabling processes involving energetic holes to be separated from processes involving energetic electrons by investigating oxidation and reduction reactions separately. Redox current densities increase by 10× under SPP excitation. The oxidation, reduction and equilibrium potentials drop by as much as 2× and split beyond a clear threshold with SPP power in correlation with the photon energy. The temperature of the working electrode under SPP excitation is monitored *in situ* and independent control experiments isolate thermal effects. Chronoamperometry measurements with on-off modulated SPPs at 600 Hz, yield a diffusion-limited rapid current response modulated at the same frequency, ruling out thermally-enhanced mass transport. Our observations are attributed to the opening of optically controlled non-equilibrium redox channels associated with energetic carrier transfer to the redox species.

**Main Text:**

Surface plasmon polaritons (SPPs) on metal surfaces have useful properties such as strong field enhancement and sub-wavelength localisation (*1*), which have long-driven vigorous interest in these waves. SPPs are damped by absorption in the metal (*2*), which limits their propagation length and lifetime. However, the absorption of SPPs creates energetic carriers (*cf.* hot electrons and holes) along the surface of the metal (*3*), which can be exploited in device applications or in catalysis (*4-7*), thereby turning what is often viewed as a drawback into a benefit. Indeed, this route for creating energetic carriers is particularly compelling, given the high efficiency with which SPPs can be excited optically. These attributes drive research on plasmonic catalysis, motivated by a desire to open reaction pathways that are inaccessible thermally, or to improve the efficiency of chemical processes by involving energetic carriers.

A metal surface supporting SPPs can double as a working electrode within an electrochemical cell. Such electrodes are constructed as a metal film on a prism in the Kretschmann configuration (*8, 9*), or as metal nanoparticles on a conductive substrate (*10, 11*). Redox reactions occurring on the surface of an electrode can be probed using SPPs localized thereon, revealing subtle details of reactions. For instance, SPPs were used to image the local current density directly on a planar working electrode (*8*). Alternatively, the optical performance of plasmonic structures can be tuned electrochemically, *e.g.*, by shifting the SPP resonance of Au nanorods via charge injection (*11*). Electrochemical cells that incorporate plasmonic structures are also of interest as multimodal chemical transducers (*9, 12*).

Recently, attention has turned to investigating the role played by energetic carriers in electrochemical reactions (*13*). Advantageously, redox currents are easily measured, proportional to reaction rates, and directly connected to experimental conditions, including plasmonic effects. Furthermore, processes involving energetic holes can be separated from processes involving energetic electrons, by investigating oxidation or reduction reactions separately.

The creation of energetic carriers via SPP absorption invariably causes the temperature of the metal to rise, with heat diffusing into the nearby reaction volume. Given that chemical reactions are temperature dependent, separating the roles of temperature and energetic carriers is not trivial but essential to proper interpretation of results (*14-18*).

Empirically and in general, a reaction rate $K$ increases exponentially with temperature following the Arrhenius law (*15, 16, 18*): $K = A \cdot \exp[-E_a/(R(T_0 + cI))]$ where $E_a$ is the molar activation energy, $R$ is the gas constant, $T_0$ is the nominal temperature, $c$ is a photothermal conversion coefficient, and $A$ is a constant depending on the reaction. It is generally a good assumption that the temperature of a plasmonic structure increases linearly with incident intensity ($T \propto cI$) if the optical response is linear. Conversely, the reaction rate should depend linearly on optical intensity if energetic carriers play a role, because their number is proportional to the number of SPP quanta absorbed, which depends on the number of incident photons (intensity). In principle, the roles of temperature and energetic carriers can be distinguished by varying the intensity and observing whether the reaction rate evolves exponentially or linearly. However, this approach has pitfalls. For instance, an exponential appears linear for small changes in argument (*cf.* Taylor series), so experiments involving small intensity changes would imply an incorrect trend.

In electrochemical systems, increasing electrode and electrolyte temperature affects the equilibrium potential and electron transfer rates to/from the redox species, and leads to convection which works to cool the system and enhance mass transport to the electrode (*19-22*). Thermal modelling including diffusion, convection and mass transport (*21*), predicts an approximately linear increase in redox currents with heating power ($i \propto P^a$, $a = 0.8$ to $1$) (rather than exponential) and a current rise time of at least 10 s upon onset of heating. Trends with temperature are not straightforward, so electrochemical cells should be thermally stabilised, the electrode temperature should be monitored, and optical variables in addition to intensity should be varied to separate the role of temperature from that of energetic carriers.

Much of the work carried out to date in plasmonic catalysis involved colloidal arrangements of Au nanoparticles illuminated at visible wavelengths (*e.g.*, $\lambda_0 = 532$ nm, $h\upsilon = 2.33$ eV) (*4 - 7, 13*). This scenario, although convenient, poses challenges. For instance, the temperature in the immediate vicinity of nanoparticles can be difficult to predict and measure due to collective effects (*16*). Also, carriers excited in Au at wavelengths above the interband threshold ($h\upsilon \sim 2$ eV) have very short lifetimes (due to electron-electron scattering at high carrier energies) (*23*).

High energy carriers ($h\upsilon > 2$ eV) are generally deemed essential to catalyse reactions via SPPs. Contrary to this broadly held view, we use here lower energy infrared photons ($\lambda_0 \sim 1350$ nm, $h\upsilon \sim 1$ eV) to excite SPPs and energetic carriers in Au. The carriers have energies at most 1 eV above $E_F$, and longer lifetimes, or longer attenuation lengths ($L_e \sim 74$ nm, $L_h \sim 55$ nm) (*24*), as the main carrier damping mechanism is electron-phonon scattering. Furthermore, we use a thin Au stripe as a SPP waveguide and working electrode (WE), which offers several advantages over colloidal nanoparticles: The WE is defined lithographically and is well-understood structurally. The temperature of the WE under SPP excitation is determined *in situ* using calibrated resistance measurements. SPPs propagate over the entire length of the WE with exclusive and complete overlap. Finally, the thickness of the WE is thinner than the excited carrier attenuation lengths in Au ($L_e$, $L_h$), enabling multiple internal carrier reflections that enhance their escape probability (*25*).

Metallic structures were fabricated on a multilayer substrate, as shown in Fig. 1, using standard nanofabrication techniques (26, 27). Fig. 1A gives a microscope image of a Au stripe designed to operate simultaneously as a SPP waveguide and WE. The image also shows a nearby Pt stripe used as a counter electrode (CE), and thick contact pads (> 200 nm) for electrical probing. The chip is immersed in a petri dish filled with the redox species in electrolyte in which a Ag/AgCl reference electrode is dipped, thereby forming a 3-electrode electrochemical cell. Glycerol was added to the electrolyte to adjust its refractive index ($n = 1.3325$, $\lambda_0 = 1312$ nm), as the solution also acts as the upper cladding of the SPP waveguide. The petri dish was mounted on a thermo-electric cooler (TEC) driven by a temperature controller using an electronic thermometer dipped in the electrolyte for feedback control. Unless stated otherwise, the bulk electrolyte temperature was maintained constant to 20 °C during all experiments. (*cf.* Materials and Methods, Fig. S1, Supplementary Materials.)

Fig. 1B illustrates the geometry of the electrodes, including their dimensions. The design of the Pt stripe (CE) is identical to that of the Au stripe (WE). Both are 35 nm thick (< $L_e$, $L_h$ in Au) and were deposited on 0.3 nm of Ti as an adhesion layer.

The geometry of the WE is constrained by its dual use as an SPP waveguide. The substrate is a multilayer stack, consisting of 15 periods of alternating layers of $SiO_2/Ta_2O_5$, as a truncated 1D photonic crystal, on a Si wafer. Bloch long-range SPPs propagate over the full length of the Au stripe with a field distribution that surrounds the stripe (*26*) – this is a key attribute of the structure as SPPs overlap exclusively and completely the WE.

The lower insets to Fig. 1A show scanning electron microscope (SEM) images of a grating coupler, as used for optical input and output coupling to Bloch long-range SPPs using perpendicular optical fibers (*26 - 28*). The experimental scheme is illustrated in Fig. 1B and briefly described in the figure caption. Advantageously, the arrangement in transmission enables optimization *in situ* of both optical alignments and the operating wavelength.

Once optical alignments were established, the laser wavelength was swept over the range 1300 to 1370 nm while the output optical power was monitored (Fig. S2, Supplementary Materials). The output power was maximum over the 1330 to 1370 nm range, which agrees well with the design wavelength of the grating couplers and waveguide (*27, 28*). Three operating wavelengths were used for the experiments: 1330, 1350 and 1370 nm.

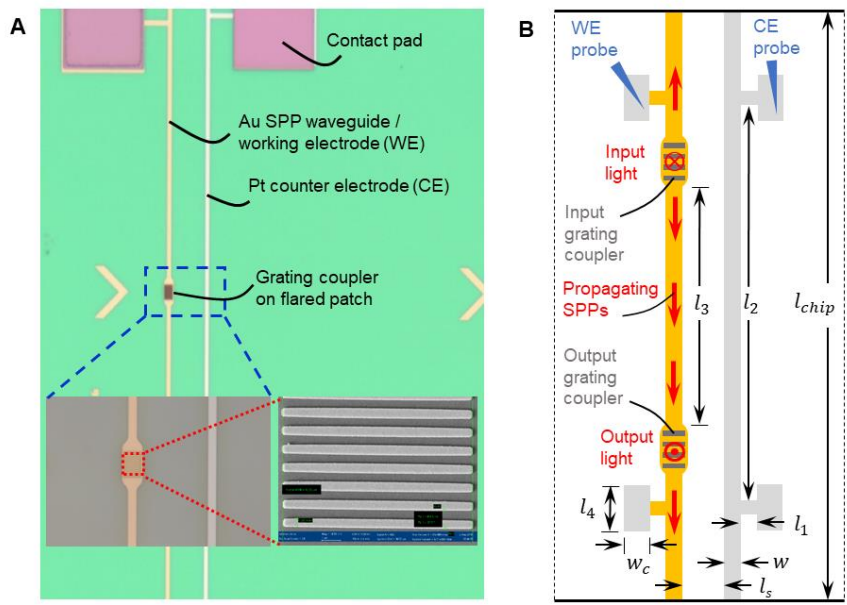

**Fig. 1. Plasmonic waveguide and electrodes.** (**A**) Microscope image of a chip bearing Au SPP waveguides / working electrodes, Pt counter electrodes, Au grating couplers, and electrical contact pads on a multilayer substrate (appears green under bright field optical microscopy). The lower insets show SEM images of a grating coupler. (**B**) Geometry of the electrodes: $l_1$ = 29 µm, $l_2$ = 2600 µm, $l_3$ = 1850 µm, $l_4$ = 250 µm, $l_{chip}$ = 3000 µm, $w_c$ = 100 µm, $w$ = 5 µm, $l_s$ = 40 µm. Experimental scheme: The WE and CE are contacted using external probes to form a 3 electrode electrochemical system with a Ag/AgCl reference electrode (not shown). Input laser light, normally-incident on the input grating coupler, excites SPPs that propagate along the WE. The output grating coupler converts the SPPs to output light emerging normally from the chip, which is captured and measured.

All experiments were carried out with 0.5 mM $K_3[Fe(CN)_6]$ + 100 mM $KNO_3$ electrolyte, at a scan rate of 100 mV/s. Applying such a potential to the system leads to the reversible reaction of potassium ferricyanide ($K_3[Fe(CN)_6]$) to potassium ferrocyanide ($K_4[Fe(CN)_6]$) as a redox couple in a 1-electron transfer process:

$$K_4[Fe(CN)_6] \leftrightarrow K_3[Fe(CN)_6] + e^-$$

Cyclic voltammetry (CV) was carried out on a Au WE, without optical illumination, providing a reference CV curve.

CV was then carried out under the same conditions but with SPPs propagating along the Au WE, as a function of incident optical power (intensity) and wavelength. The measured CV curves, shown in Fig. 2A for $\lambda_0$ = 1350 nm, change dramatically as the optical power increases (the output optical power was monitored in all experiments).

The current increases significantly with optical power compared to the reference case (no illumination). Fig. 2B plots the peak oxidation and reduction currents *vs*. the output optical power, for the three wavelengths of interest, revealing an increase of ~10× in both over the power range investigated. The reduction current increases more than the oxidation current. A clear threshold is observed in Fig. 2B at an output power of ~0.1 µW, about which different linear trends are evident (linear models fitted for each segment at $\lambda_0$ = 1350 nm are plotted as the solid black lines).

The potentials corresponding to the peak redox currents also change significantly with optical power, as shown in Fig. 2C. The reduction potential decreases by ~2×, the oxidation potential by ~1.3×, and the equilibrium potential (mean of the redox potentials) by ~1.7×. Again, a clear threshold is observed at an output power of ~0.1 µW. Beyond threshold, the redox potentials decrease commensurately with increasing photon energy: $\Delta(h\upsilon)$ = 13.8 meV separates $\lambda_0$ = 1330 from 1350 nm, and $\Delta(h\upsilon)$ = 13.4 meV separates $\lambda_0$ = 1350 from 1370 nm – vertical blue bars of 14 mV on Fig. 2C illustrate this point.

The resistance of the WE was measured *in situ* as a function of optical power, and compared to calibrated resistances in order to determine its temperature (Fig. S3, Supplementary Materials). The temperature of the WE, added to Figs. 2B and 2C as the top horizontal scale, spans about 30 ºC.

Various control experiments were carried out *in situ* under identical experimental conditions.

Control experiments without the redox species, produced featureless and noisy CV curves, whether the illumination was on or off.

Control experiments with the laser on while misaligning the input optical fibre (in various ways) were carried out. The CV curves would always return to the reference case (no illumination) as soon as coupling to the input grating was lost, confirming that the excitation of SPPs on the WE was essential to the changes observed in Fig. 2.

Temperature control experiments without illumination were carried out using the TEC placed under the petri dish to cool then heat the entire cell (electrodes and electrolyte) in a controlled manner (Fig. S4, Supplementary Materials). CV curves obtained at different temperatures, ranging from 10 to 40 °C, show changes in peak redox currents of about 30%, following linear trends with temperature of fitted slopes ±4 nA/°C, consistent with thermally-induced mass-transport effects (*21*). The reduction and oxidation potentials change by 10% to 20%. The temperature dependence of the equilibrium potential was fit to a linear model yielding a slope of -1.4 mV/°C, in good agreement with thermally induced shifts reported in the literature for the ferricyanide/ferrocyanide system (*19, 29-31*). Linear thermal trends based on these measurements were added to Figs. 2B and 2C as the dashed blue lines.

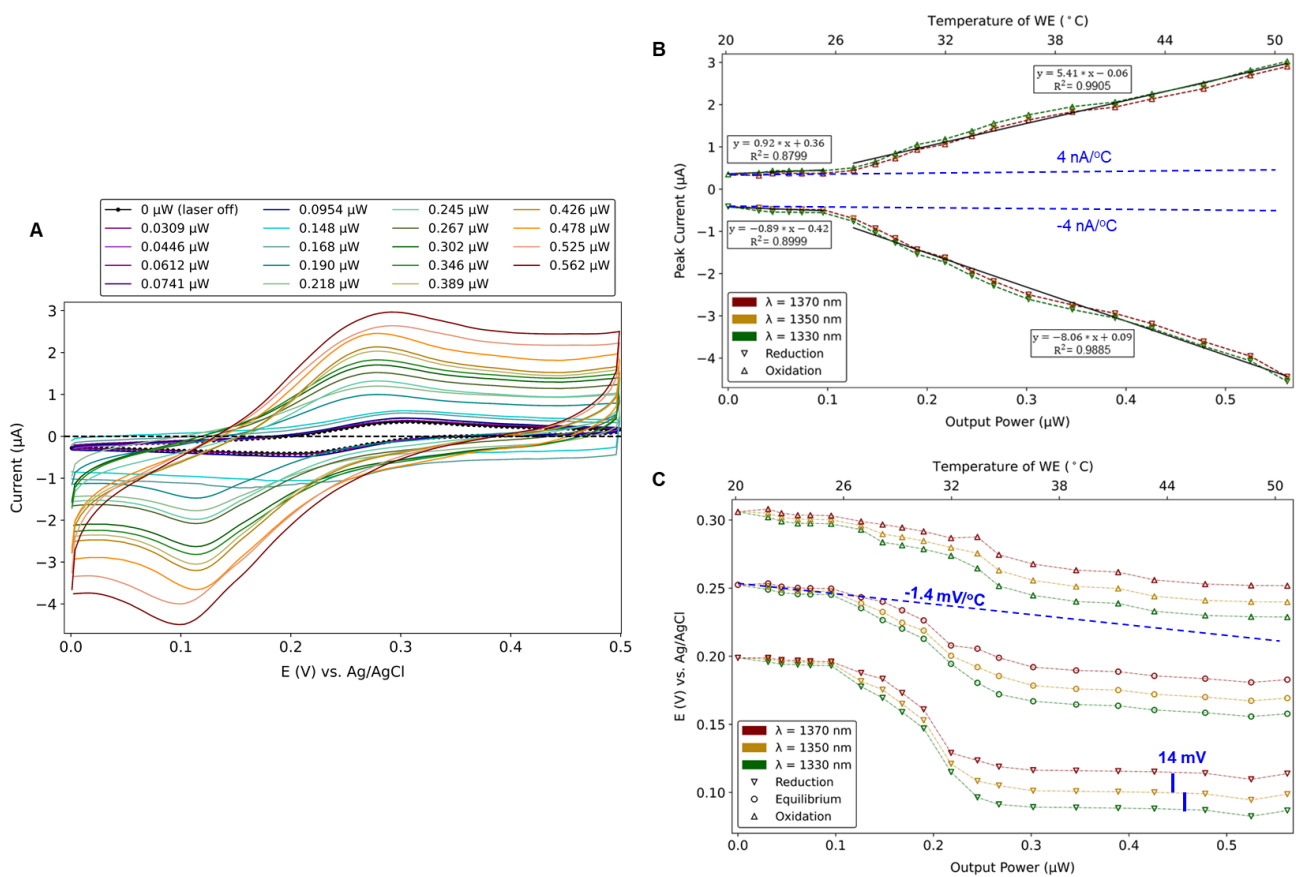

**Fig. 2. Cyclic voltammetry under optical illumination.** (**A**) CV curves obtained on a Au WE, in 0.5 mM $K_3[Fe(CN)_6]$ + 100 mM $KNO_3$ electrolyte, at a scan rate of 100 mV/s, for increasing optical power (legend) at $\lambda_0$ = 1350 nm. The incident optical power ranged from 0 to 6.3 mW. The reference CV curve (laser off) is plotted in black dots. (**B**) Redox current peaks, and (**C**) potentials *vs*. optical power, from CV curves measured at $\lambda_0$ = 1330, 1350 and 1370 nm. Linear models of the peak redox currents at $\lambda_0$ = 1350 nm are plotted as the solid black lines in Part (B) (slopes have units of A/W). The solid vertical blue bars in Part (C) measure 14 mV, corresponding to the approximate photon energy separating the three optical wavelengths (~14 eV). Linear thermal trends, measured independently, are added as the blue dashed lines to Parts (B) and (C).

Thermally induced mass transport effects were ruled out by taking chronoamperometry measurements while the laser was internally modulated on/off at a frequency of 600 Hz (period of 1.67 ms). The inset to Fig. 3A shows the time response of the modulated laser power measured using a photoreceiver. The modulation frequency of 600 Hz is low enough to enable our potentiostat to reliably acquire several current samples (8) within a modulation period (1.67 ms), yet high enough to preclude thermally induced mass transport, which occurs on timescales of the order of seconds (*21*).

The chronoamperometric response is shown in Fig. 3A. A forward step potential of 450 mV (*vs*. Ag/AgCl) was applied from a null potential to induce oxidation of the redox species at the Au WE, while the latter supports on/off modulated propagating SPPs. The incident laser power was modulated from 0 (off-state) to 6.3 mW (on-state), corresponding to the incident power extrema investigated in Fig. 2. Thus, the step potential of 450 mV is well above the oxidation potential at both powers (Fig. 2), which ensures that the current is diffusion limited.

The current response of Fig. 3A was plotted *vs.* $t^{-1/2}$ for the purpose of identifying the region where it decays following the Cottrell equation, indicating diffusion-limited conditions. Fig. 3B plots a zoom-in of the chronoamperometric response in the Cottrell region, at the location of the blue arrow sketched in Fig. 3A, revealing a periodic response tracking in time the laser modulation shown inset to Fig. 3A. The current during the laser off-state is below the noise floor of our instrument, but the current during the on-state has a high signal-to-noise ratio. Recalling that thermally induced mass transport occurs over timescales of seconds rules out this process as being responsible for the increase in current.

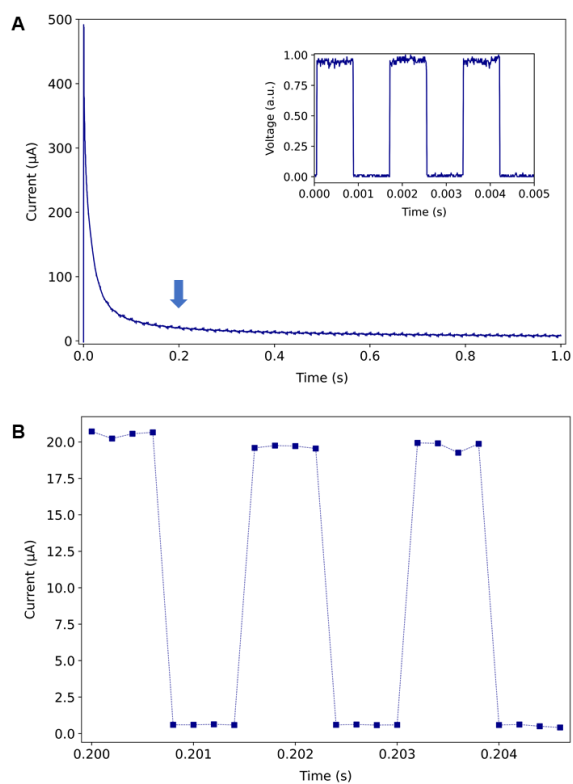

**Fig. 3. Chronoamperometry under optical modulation.** (**A**) Chronoamperometric response on a Au WE to a 450 mV potential step, in 0.5 mM $K_3[Fe(CN)_6]$ + 100 mM $KNO_3$ electrolyte, for on/off laser modulation. The inset shows the on/off laser modulation response measured using a photoreceiver (modulation frequency of 600 Hz, period of 1.67 ms). (**B**) Zoom-in of the chronoamperometric response in the Cottrell region, at the location of the blue arrow sketched in Part (A), resolved as 8 current samples per period.

We conclude that the enhancements observed in the redox currents (Figs. 2 and 3) are due to energetic carriers created along the Au WE as SPPs propagate and are absorbed therein. Energetic electrons transfer more easily from the WE to the redox species enhancing the reduction current, and energetic holes transfer more easily to enhance the oxidation current – both are required to explain the results observed in Fig. 2B. Both trends are also linear with optical power, as expected for an effect based on energetic carriers. Thermal effects on the redox currents are also linear with temperature, but follow much weaker trends as noted from the measured thermal trendlines added to Fig. 2B (dashed blue).

The creation of energetic carriers also explains the drop in the oxidation and reduction potentials observed in Fig. 2C, as lower potentials are needed to drive energetic carrier transfer to the redox species. The drop goes well-beyond the thermal trendline for the equilibrium potential added to Fig. 2C (dashed blue).

The power threshold behaviour observed in Figs. 2B and 2C indicates the opening of redox channels associated with energetic carrier transfer, as these channels overcome noise in the system and the equilibrium redox currents to become dominant. The redox potentials beyond threshold decrease commensurately with increasing photon (SPP) energy as carriers become more energetic – the potentials are controlled optically via the photon energy.

Finally, infrared SPPs (photons) of energy $h\upsilon \sim 1$ eV create electrons and holes in Au that are sufficiently energetic and long-lived to significantly affect electrochemical reactions.

**Acknowledgements:**

The authors are grateful to Dr. Li Sun, Pine Research for assistance with the electrochemical measurements. **Funding:** Financial support provided by the Natural Sciences and Engineering Research Council of Canada is gratefully acknowledged under grant 210396. **Author contributions:** Z. H. carried out the optical and electrochemical experiments, contributed to measurement interpretation, and prepared the first draft of the manuscript. O. K. assisted with the measurements and interpretation. H. N and A. O. fabricated the waveguide structures tested. E. A. B. assisted with the experimental techniques and the interpretation of the measurements. P. B. directed the research. All Authors contributed to writing the manuscript. **Competing interests:** The authors declare no competing interests. **Data and materials availability:** All data is available in the manuscript or the supplementary materials.

**Supplementary Materials:**

Materials and Methods
Figs. S1 – S4

Supplementary Materials for

**Surface plasmons on a Au waveguide electrode open new redox channels associated with the transfer of energetic carriers**

Zohreh Hirbodvash,[1,2] Oleksiy Krupin,[2] Howard Northfield,[2] Anthony Olivieri,[2] Elena A. Baranova,[3,4] and Pierre Berini[1,2,5,*]

[1]Dept. of Physics, University of Ottawa, 150 Louis Pasteur, Ottawa, Ontario, K1N 6N5, Canada
[2]Center for research in Photonics, University of Ottawa, 25 Templeton St., Ottawa, Ontario, K1N 6N5, Canada
[3]Department of Chemical and Biological Engineering, University of Ottawa, 161 Louis-Pasteur, Ottawa, ON K1N 6N5, Canada
[4]Centre for Catalysis Research and Innovation, University of Ottawa, 161 Louis-Pasteur, Ottawa, ON K1N 6N5, Canada
[5]School of Electrical Engineering and Computer Science, University of Ottawa, 800 King Edward Ave., Ottawa, Ontario, K1N 6N5, Canada
*Corresponding Author; berini@eecs.uottawa.ca

**This file includes:**

    Materials and Methods
    Figures S1 to S4

**Materials and Methods**

*Preparation of the waveguides and electrodes on chip*

Chips incorporating Au stripes with contact pads and grating couplers, and Pt stripes with contact pads were fabricated as described in previous work (*26,27*) but without upper cladding or lid. Chips without an upper cladding enable direct probing of pads connected to the electrodes and facilitate alignment of the optical fibers used to couple to the gratings on chip. The upper cladding consisted of the electrolyte solution, prepared as described below. A Au stripe was used simultaneously as an SPP waveguide and working electrode, whereas a nearby Pt stripe was used as a counter electrode.

The chips were cleaned manually using a swab dipped in acetone to remove protective photoresist (SPR-220) from the surface. Then, a chip was placed into a large glass vial in acetone, and the vial was placed in an ultrasonic bath for 5 minutes. Next, the chip was promptly rinsed with 3-4 isopropyl alcohol (IPA) via pipetting, then with deionized water. Pressurized $N_2$ gas was used to dry the chip. Finally, the chip was placed in a UV/Ozone chamber for 30 minutes with the UV lamp on followed by 30 minutes with the UV lamp light off, as a final cleaning step.

Working and counter electrodes were then selected and electrically burned-in before use by injecting a current (ramp function) along an electrode structure until its resistance stabilised. Burn-in induces grain reorganization through annealing, which stabilizes the electrodes for use in an electrochemical experiment (*32*).

The chip was affixed to the bottom of a petri dish and electrochemical, optical or resistance measurements were carried out as required.

*Experimental set-up*

The experimental setup, constructed on a floating optical table, includes a tunable laser (8164A, Agilent) working over the wavelength range from 1270 to 1370 nm. A cleaved bow-tie style polarization-maintaining single-mode optical fiber (PM-SMF) with a 6.6 µm core diameter and a multi-mode fiber (MMF) with a core diameter of 200 µm were aligned perpendicularly to the input and output grating couplers, respectively. 90° metallic holders mounted on multi-axis micro-positioners were used to hold the fibers, while ensuring that TM-polarized light was incident on the input grating. A power meter (PM 100USB, Thoarlabs) was used to monitor the power emerging from the MMF coupled to the output grating. Following the scheme illustrated in Fig. 1B of the main text, the input PM-SMF was aligned perpendicularly to the input grating coupler such that laser light is incident thereon, exciting Bloch long-range SPPs along the Au stripe in both longitudinal directions. Propagating SPPs are outcoupled by the output grating into light captured by the MMF connected to the power meter. This arrangement in transmission enables optimization of both optical alignments, as well as the operating wavelength.

Two tungsten needles, attached to the arms of two micro-positioners, were used to probe pads at the end of a Pt stripe and a Au stripe, following the scheme illustrated in Fig. 1B of the main text. The needles were coated with poly(methyl methacrylate) except for their tip to ensure no interaction with the electrolyte.

A bipotentiostat (WaveDriver 20, basic bundles, Pine Research Instrumentation) and a Ag/AgCl reference electrode (double junction PH combination, glass body, BNC connector, Sigma-Aldrich) were used to perform the electrochemical measurements. The chip was immersed in electrolyte, the reference electrode was dipped nearby, and the needles were used to probe the pads of the working and counter electrodes on chip to connect them to the bipotentiostat in a 3 electrode configuration. Cyclic voltammetry and chronoamperometry routines supplied by the manufacturer of our bipotentiostat were used to carry out the measurements.

The petri dish housing the chip and electrochemical cell was placed on a thermo-electric cooler (TEC). The interface between the petri dish and TEC was filled with silicon paste to ensure high thermal conductivity between these parts. A clean electronic thermometer was dipped in the electrolyte and used with a temperature controller (operating in a closed-loop feedback control algorithm) to control the temperature of the cell via the TEC, over the range from 10 to 40 ºC. Sufficient time was allowed to lapse whenever the temperature was changed to allow the cell and its contents (electrolyte, chip, electrodes) to reach thermal equilibrium before any measurements were undertaken.

An electronic source meter (2400, Keithley) along with an extra pair of probe needles and micro-positioners were used to measure stripe resistances.

An alignment microscope of long working distance was used to align the fibres and probes to the chip.

Schematics of our experimental arrangement and set-up are given in Fig. S1.

*Redox species and electrolyte / upper cladding*

Potassium ferricyanide ($K_3[Fe(CN)_6]$) and potassium nitrate ($KNO_3$) were used as supplied (Sigma-Aldrich) as the redox species and supporting electrolyte, respectively.

Glycerol (Sigma-Aldrich) was used as supplied and added to the electrolyte (0.2928 g of glycerol per 20 ml of electrolyte) to adjust the refractive index of the solution, because it also acts as the upper cladding of the SPP waveguide. The refractive index of the solution was adjusted to $n = 1.3325$ ($\lambda_0 = 1312$ nm), as measured using a prism coupler (Metricon), such that the Bloch long-range SPP could propagate (*26*). CV measurements with and without glycerol were identical over the range of potentials of interest (no glycerol oxidation was observed over the potential window of 0 to 0.5 V *vs.* Ag/AgCl used in our study).

**Supplementary Figures**

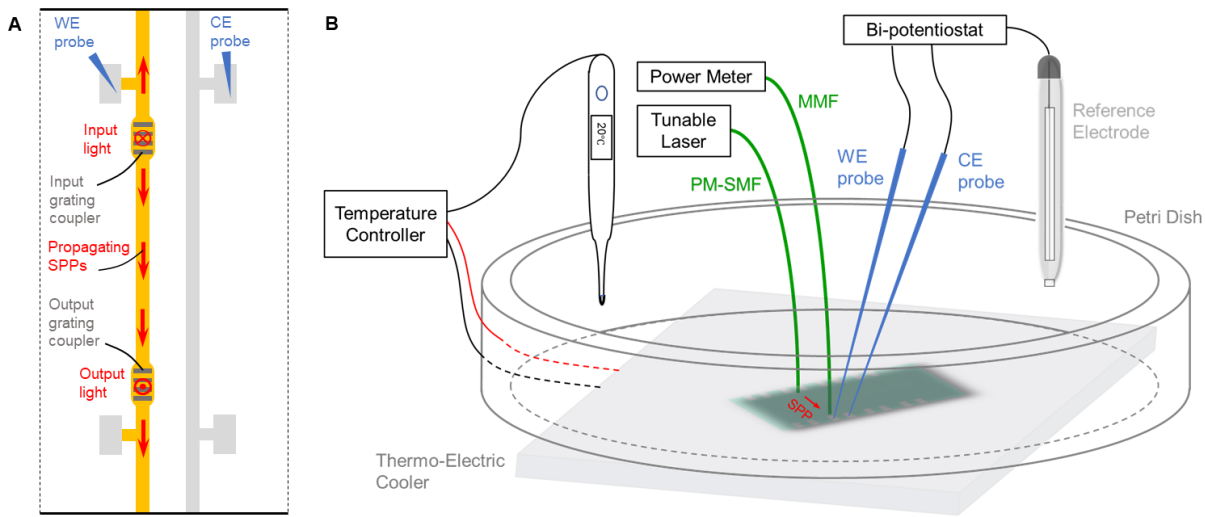

**Fig. S1. Experimental set-up**. (**A**) Sketch of chip and (**B**) block diagram of experimental set-up. PM-SMF: polarization-maintaining single-mode fibre; MMF: multi-mode fibre; WE: working electrode; CE: counter electrode.

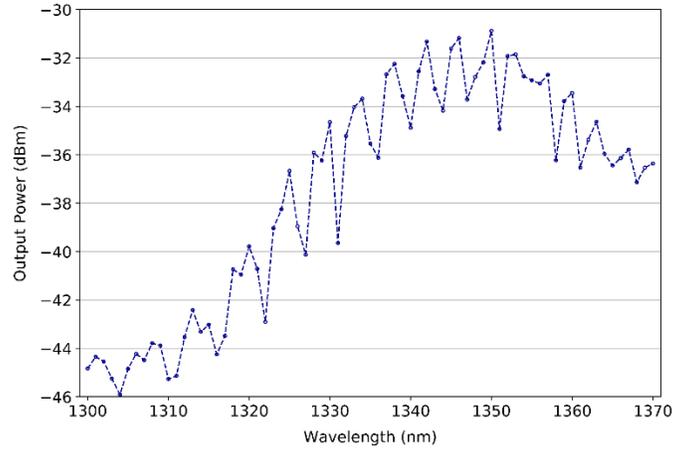

**Fig. S2. Wavelength response of grating-coupled waveguide / working electrode.** Wavelength response of grating couplers, separated by a waveguide segment of length $l_3 = 1850$ µm (Fig. 1B, main text). The incident optical power was set to 8 dBm. The combined grating coupling losses are ~20 dB, and the propagation loss of the Bloch LRSPP along the stripe is ~11 dB/mm (*27*).

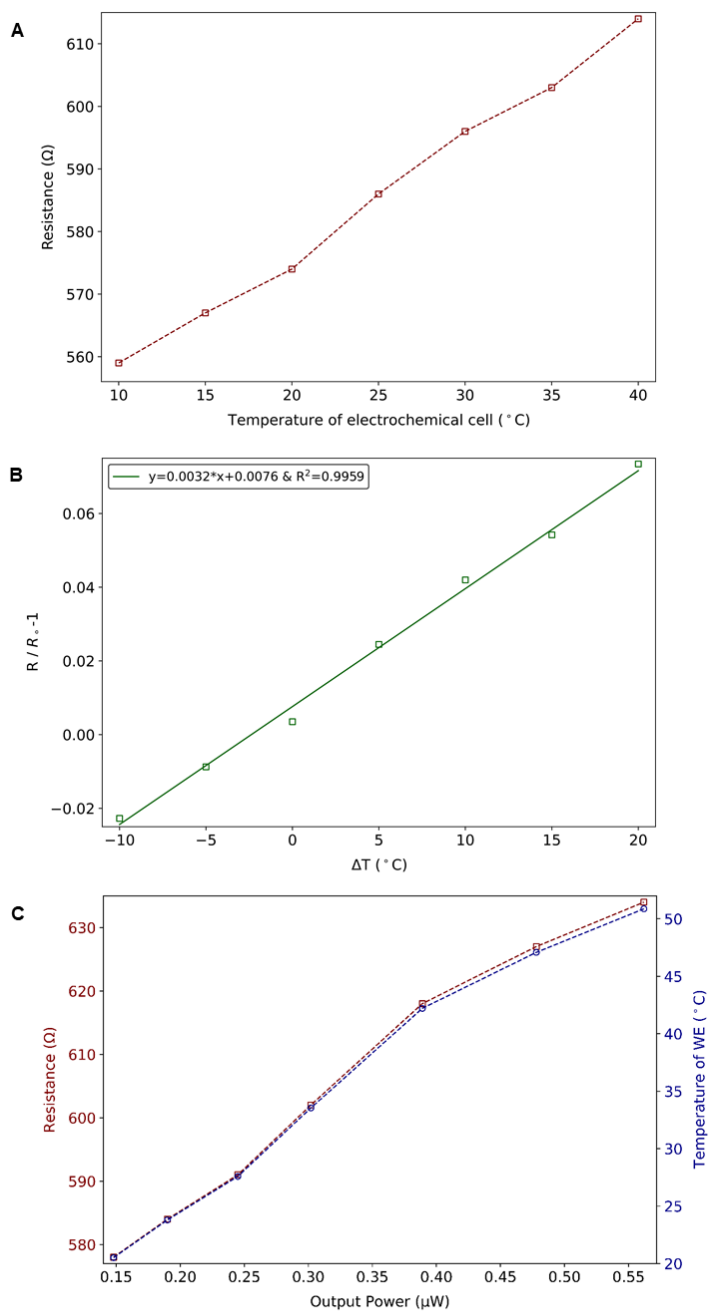

**Fig. S3. Resistance of WE as a function of temperature and optical power**. (**A**) Resistance of a Au WE measured *in situ vs.* temperature of the electrochemical cell controlled using the TEC. (**B**) Results of Part (A) re-plotted using $R = R_0[1+\alpha(T - T_0)]$ such that the temperature coefficient of resistivity ($\alpha$) appears as the slope; $R_0$ is the resistance at the reference temperature of $T_0 = 20$ °C, $R$ is the resistance at temperature $T$, and $\Delta T = T - T_0$. The best fit linear model given in the legend yields $\alpha = 3.2 \times 10^{-3}$ °C$^{-1}$, in excellent agreement with the literature for bulk Au (*33*). (**C**) Resistance of Au WE measured *in situ vs.* increasing output optical power, with the electrochemical cell maintained at 20 °C. The corresponding temperature of the WE is plotted on the right axis, deduced using the relation $R = R_0[1+\alpha(T - T_0)]$ with $\alpha$ as measured in Part (B).

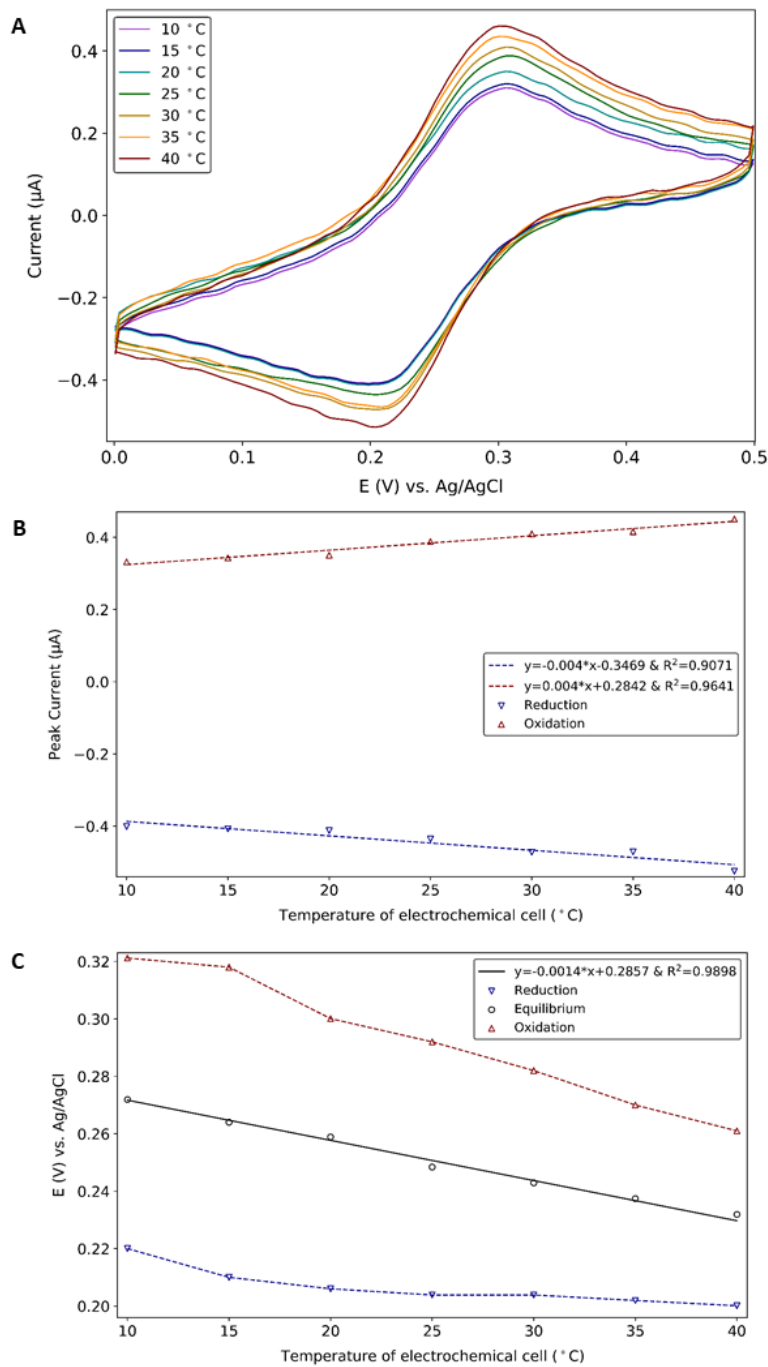

**Fig. S4. Cyclic voltammetry at various cell temperatures.** (**A**) CV curves obtained on a Au WE, in 0.5 mM $K_3[Fe(CN)_6]$ + 100 mM $KNO_3$ electrolyte, at a scan rate of 100 mV/s, as the temperature of the electrochemical cell was varied using the TEC from 10 to 40 °C (legend). (**B**) Redox current peaks, and (**C**) potentials *vs.* temperature, obtained from the CV curves of Part (A). Linear models fitted to the peak redox currents are plotted as the dashed lines in Part (B) and given in the legend (slopes have units of µA/°C). A linear model fitted to the equilibrium potential (mean of redox potentials) is plotted as the solid black line in Part (C) and given in the legend (slope has units of V/°C). These linear models are transposed to Fig. 2 of the main text (*cf.* dashed blue lines).

21